\def\Slash#1{\not\!\!#1}

\documentclass{webofc}
\usepackage[varg]{txfonts}   
\woctitle{Analytical relation between quark
confinement and chiral symmetry breaking in odd-number lattice QCD}
\begin{document}
\title{Analytical relation between quark 
confinement and chiral symmetry breaking in odd-number lattice QCD}
%
%

\author{Hideo Suganuma\inst{1}\fnsep\thanks{\email{suganuma@scphys.kyoto-u.ac.jp}} \and
        Takahiro M. Doi \inst{1}\fnsep
\and
        Takumi Iritani \inst{2}\fnsep
}

\institute{
        Department of Physics \& Division of Physics and Astronomy, 
Graduate School of Science, \\
Kyoto University, 
Kitashirakawaoiwake, Sakyo, Kyoto 606-8502, Japan
\and
High Energy Accelerator Research Organization (KEK), 
Tsukuba, Ibaraki 305-0801, Japan
          }

\abstract{%
To clarify the relation between confinement and 
chiral symmetry breaking in QCD, 
we consider a temporally odd-number lattice, 
with the temporal lattice size $N_t$ being odd. 
We here use an ordinary square lattice with 
the normal (nontwisted) periodic boundary condition for link-variables 
in the temporal direction. 
By considering ${\rm Tr} (\hat{U}_4\hat{\Slash{D}}^{N_t-1})$, 
we analytically derive a gauge-invariant relation between the Polyakov loop 
$\langle L_P \rangle$ and the Dirac eigenvalues $\lambda_n$ in QCD, i.e., 
$\langle L_P \rangle \propto 
\sum_n \lambda_n^{N_t -1} \langle n|\hat U_4|n \rangle$, 
which is a Dirac spectral representation of 
the Polyakov loop in terms of Dirac eigenmodes $|n\rangle$.
Owing to the factor $\lambda_n^{N_t -1}$ in the Dirac spectral sum, 
this relation generally indicates fairly small contribution of 
low-lying Dirac modes to the Polyakov loop, 
while the low-lying Dirac modes are essential for chiral symmetry breaking. 
Also in lattice QCD calculations in both confined and deconfined phases, 
we numerically confirm the analytical relation, 
non-zero finiteness of $\langle n|\hat U_4|n \rangle$ for each Dirac mode, 
and negligibly small contribution from low-lying Dirac modes 
to the Polyakov loop, i.e., 
the Polyakov loop is almost unchanged 
even by removing low-lying Dirac-mode contribution from 
the QCD vacuum generated by lattice QCD simulations. 
We thus conclude that low-lying Dirac modes 
are not essential modes for confinement, 
which indicates no direct one-to-one correspondence between 
confinement and chiral symmetry breaking in QCD.
}
\maketitle

\section{Introduction}

Color confinement and 
spontaneous chiral-symmetry breaking \cite{NJL61} 
are the two outstanding nonperturbative phenomena in 
quantum chromodynamics (QCD), and they have been 
studied as important unsolved subjects in theoretical physics.
In particular, to clarify their precise relation is one of 
the challenging important issues  
\cite{SST95,M95W95,G06BGH07,SWL08,S1112,GIS12,IS13,SDI13,DSI13},
and their relation is not yet clarified directly from QCD.

For quark confinement, 
the Polyakov loop $\langle L_P \rangle$ is 
one of the typical order parameters, and 
relates to the single-quark free energy $E_q$ as 
$\langle L_P \rangle \propto e^{-E_q/T}$ 
at temperature $T$. 
The Polyakov loop is the order parameter 
of spontaneous breaking of the $Z_{N_c}$ center symmetry 
in QCD \cite{Rothe12}.
Also, its fluctuation is recently found to be important 
in the QCD phase transition \cite{LFKRS13}.

For spontaneous chiral-symmetry breaking, 
the standard order parameter 
is the quark condensate $\langle \bar qq \rangle$, 
and low-lying Dirac modes are known to be essential, 
as the Banks-Casher relation shows \cite{BC80}.

There are several circumstantial evidence 
of correlation between 
confinement and chiral symmetry breaking. 
For example, lattice QCD simulations have suggested 
almost coincidence between deconfinement and chiral-restoration temperatures 
\cite{Rothe12,K02}, although slight difference of about 25MeV between them 
is pointed out in some recent lattice QCD studies \cite{AFKS06}.
Their correlation is also suggested 
in terms of QCD-monopoles \cite{SST95,M95W95}, 
which topologically appear in QCD in the maximally Abelian gauge 
\cite{KSW87,SNW94,AS99,IS9900}, 
leading to the dual-superconductor picture \cite{N74tH81}. 
As schematically shown in Fig.1, 
confinement and chiral symmetry breaking are 
simultaneously lost in lattice QCD,
by removing the monopoles from the QCD vacuum \cite{M95W95}.
This means a crucial role of QCD-monopoles 
to both confinement and chiral symmetry breaking, so that 
these two phenomena seem to be related through the monopole.
As a possibility, however, to remove the monopoles 
may be ``too fatal'' for nonperturbative properties. 
If this is the case, nonperturbative phenomena 
are simultaneously lost by their removal. 

In fact, {\it if only the relevant ingredient of 
chiral symmetry breaking is carefully removed from the QCD vacuum, 
how will be quark confinement?}

To obtain the answer, we perform a direct investigation between 
confinement and chiral symmetry breaking, 
using the Dirac-mode expansion and projection \cite{S1112,GIS12,IS13}.

\begin{figure}[h]
\begin{center}
\includegraphics[scale=0.3]{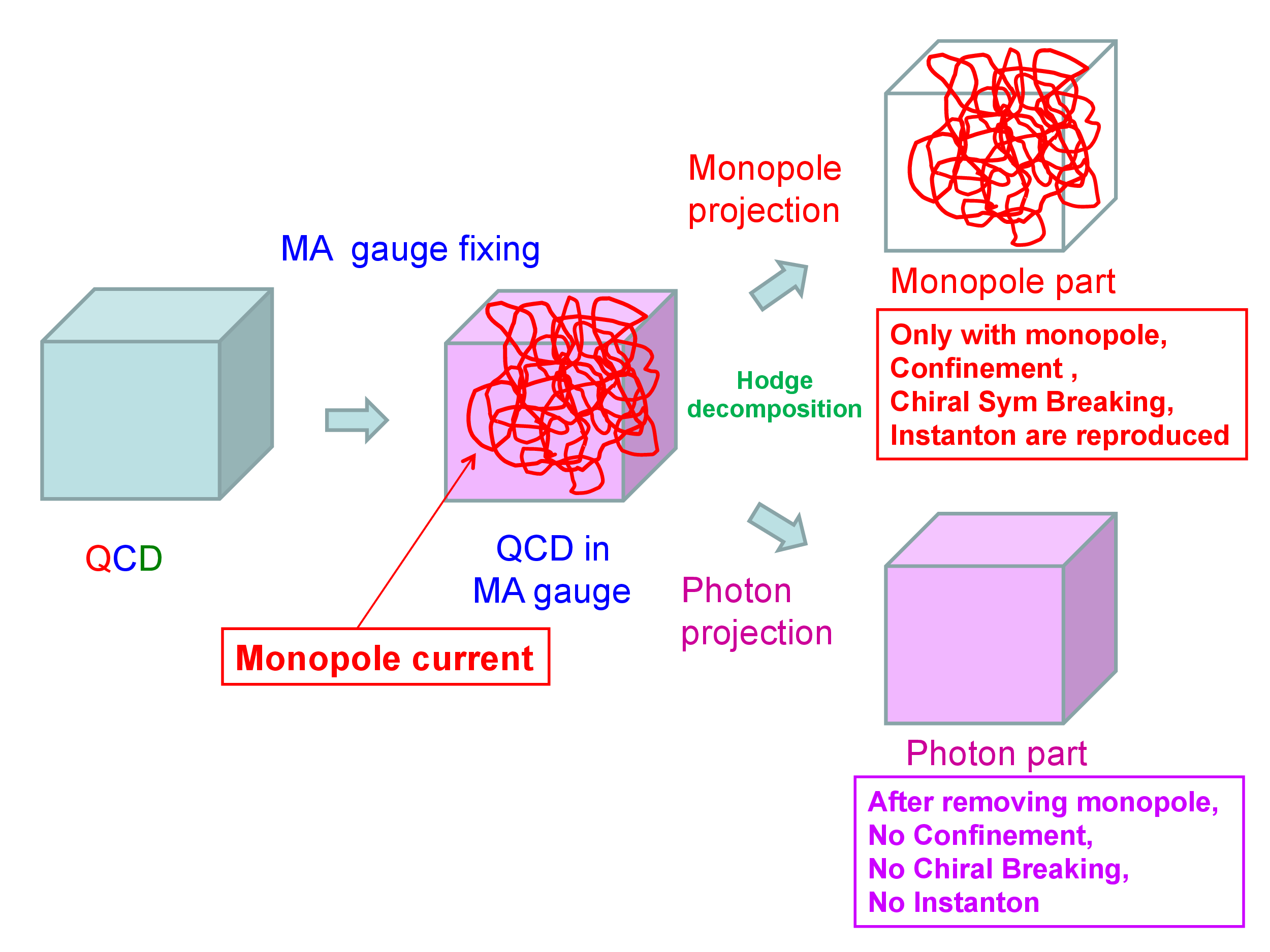}
\hspace{0.4cm} \includegraphics[scale=0.25]{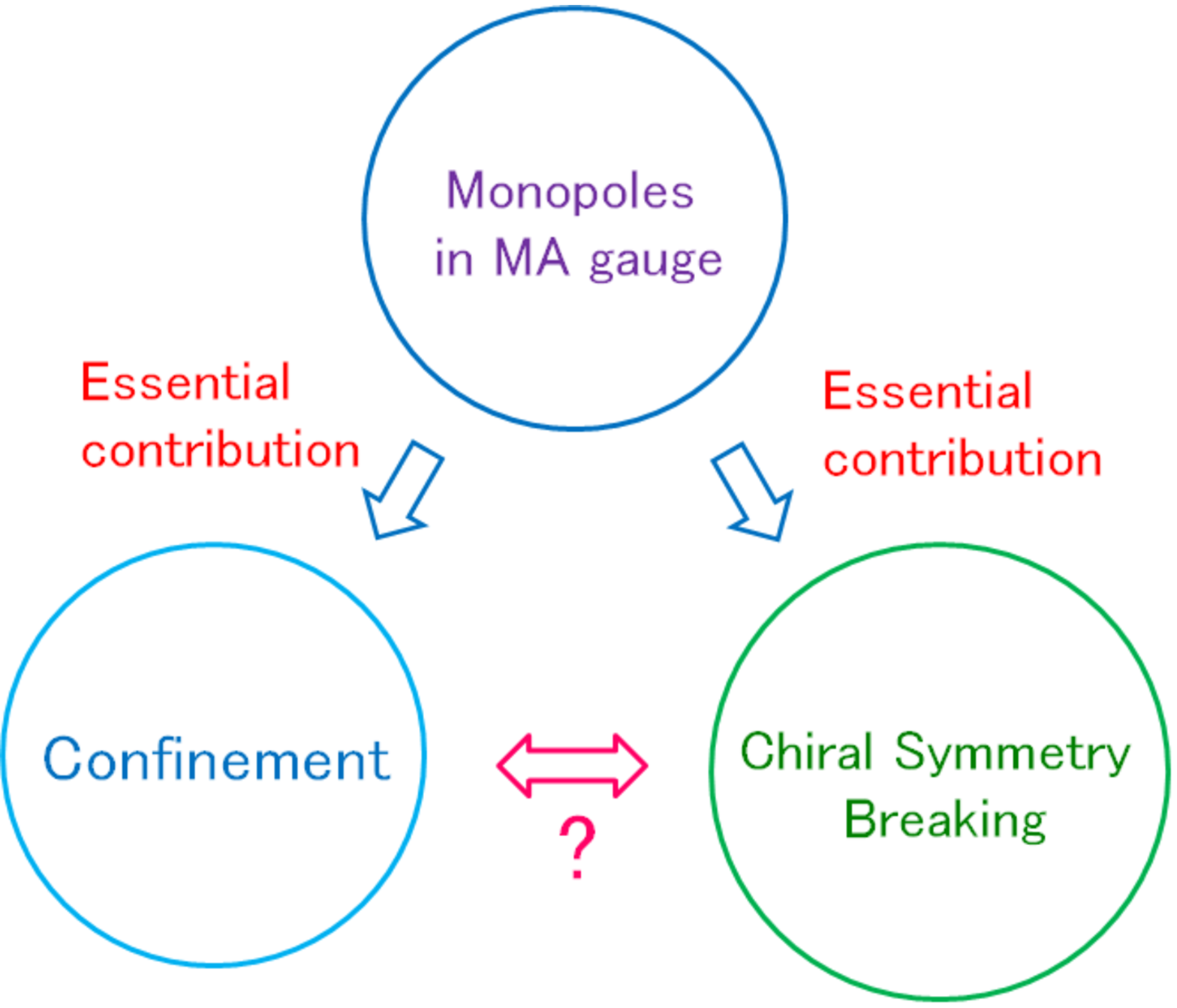}
\vspace{-0.25cm}
\caption{
The role of QCD-monopoles to nonperturbative QCD. 
In the MA gauge, QCD becomes Abelian-like owing to a 
large effective mass ($\simeq$ 1GeV) of off-diagonal gluons \cite{AS99}, 
and QCD-monopoles topologically appear as 
$\Pi_2({\rm SU}(N_c)/{\rm U}(1)^{N_c-1})={\bf Z}^{N_c-1}$ \cite{KSW87,IS9900}. 
By the Hodge decomposition, the QCD vacuum can be divided into 
the monopole part and the photon part. 
The monopole part has confinement \cite{SNW94},
chiral symmetry breaking \cite{M95W95} and instantons \cite{STSM95},
while the photon part does not have all of them, 
as lattice QCD studies show. 
In spite of the essential role of monopoles, 
the direct relation of confinement and chiral symmetry breaking 
is still unclear.
}
\end{center}
\end{figure}

In our previous works, 
we investigated the relation between 
confinement and chiral symmetry breaking in more direct manner 
\cite{S1112,GIS12,IS13} 
by analyzing confinement in terms of Dirac eigenmodes in QCD,
because of the essential role of low-lying Dirac modes 
for chiral symmetry breaking \cite{BC80}.
Using completeness of the Dirac-mode basis, 
we proposed ``Dirac-mode expansion'' and 
``Dirac-mode projection'' to a restricted Dirac-mode space, 
and investigated the role of low-lying Dirac modes 
to confinement in SU(3) lattice QCD 
\cite{S1112,GIS12,IS13}.
As the remarkable facts, 
even by the removal of the coupling to low-lying Dirac modes, 
we numerically obtained the following lattice-QCD results: 
\begin{itemize}
\item The Wilson loop obeys the area law, 
which means a linear quark confinement potential \cite{S1112,GIS12}. 
\item The slope parameter, i.e., the string tension 
or the confining force, is almost unchanged \cite{S1112,GIS12}.
\item The Polyakov loop remains to be almost zero, 
which means $Z_3$-unbroken confinement phase \cite{IS13}.
\end{itemize}
Thus, quark confinement properties are almost kept 
even in the absence of low-lying Dirac modes. 
(Also, ``hadrons'' appear without low-lying Dirac modes 
 \cite{LS11}, suggesting survival of confinement.)
In our studies, we just consider 
the mathematical expansion by eigenmodes of 
the Dirac operator $\Slash D=\gamma_\mu D_\mu$. 
For eigenmode expansions, one can deal with any (anti)hermite operator, 
e.g., $D^2=D_\mu D_\mu$. 
However, to link with chiral symmetry breaking, 
we adopt $\Slash D$ and the expansion by its eigenmodes.

In this study, we consider temporally odd-number lattice QCD, 
where the temporal lattice size is odd-number, 
and derive an analytical relation between 
the Polyakov loop and the Dirac modes. Based on the analytical formula, 
we discuss the relation between confinement and chiral symmetry breaking.

\section{Lattice QCD formalism}

In this section, we exhibit the mathematical conditions of 
lattice QCD formalism adopted in this study. 
We use an ordinary square lattice with spacing $a$ and 
size $N_s^3 \times N_t$, and impose 
the normal (nontwisted) periodic boundary condition 
for the link-variable $U_\mu(s)={\rm e}^{iagA_\mu(s)}$ 
in the temporal direction.
($A_\mu(s)$ is the gluon field, $g$ the gauge coupling, and $s$ the site.)
This temporal periodicity is physically required at finite temperature.
As the gauge group, we here take SU($N_c$) with $N_c$ being the color number. 
However, arbitrary gauge group $G$ can be taken 
for most arguments in this paper. 

\subsection{Dirac operator, Dirac eigenvalues and Dirac modes in lattice QCD}

On lattices, the Dirac operator 
$\Slash D = \gamma_\mu D_\mu$ is written with 
$U_\mu(s)={\rm e}^{iagA_\mu(s)}$ 
and $U_{-\mu}(s)\equiv U^\dagger_\mu(s-\hat \mu)$ as
\begin{eqnarray}
 \Slash{D}_{s,s'} 
 \equiv \frac{1}{2a} \sum_{\mu=1}^4 \gamma_\mu 
\left[ U_\mu(s) \delta_{s+\hat{\mu},s'}
 - U_{-\mu}(s) \delta_{s-\hat{\mu},s'} \right].
\end{eqnarray}
Here, $\hat\mu$ is $\mu$-directed vector with $|\hat \mu|=a$.
Adopting hermite $\gamma$-matrices as $\gamma_\mu^\dagger=\gamma_\mu$, 
the Dirac operator $\Slash D$ is anti-hermite and satisfies 
$\Slash D_{s',s}^\dagger=-\Slash D_{s,s'}$.
We introduce the normalized Dirac eigen-state $|n \rangle$ as 
\begin{eqnarray}
\Slash D |n\rangle =i\lambda_n |n \rangle, \qquad
\langle m|n\rangle=\delta_{mn}, 
\end{eqnarray}
with the Dirac eigenvalue $i\lambda_n$ ($\lambda_n \in {\bf R}$).
Due to $\{\gamma_5,\Slash D\}=0$, the state 
$\gamma_5 |n\rangle$ is also an eigen-state of $\Slash D$ with the 
eigenvalue $-i\lambda_n$. 
Here, the Dirac eigen-state $|n \rangle$ 
satisfies the completeness of 
\begin{eqnarray}
\sum_n |n \rangle \langle n|=1.
\end{eqnarray}
For the Dirac eigenfunction $\psi_n(s)\equiv\langle s|n \rangle$, 
the explicit form of the Dirac eigenvalue equation \\
$\Slash D \psi_n(s)=i\lambda_n \psi_n(s)$ is expressed in lattice QCD as 
\begin{eqnarray}
\frac{1}{2a} \sum_{\mu=1}^4 \gamma_\mu
[U_\mu(s)\psi_n(s+\hat \mu)-U_{-\mu}(s)\psi_n(s-\hat \mu)]
=i\lambda_n \psi_n(s).
\end{eqnarray}
The Dirac eigenfunction $\psi_n(s)$ can be 
numerically obtained in lattice QCD, besides a phase factor. 
By the gauge transformation of 
$U_\mu(s) \rightarrow V(s) U_\mu(s) V^\dagger (s+\hat\mu)$, 
$\psi_n(s)$ is gauge-transformed as 
\begin{eqnarray}
\psi_n(s)\rightarrow V(s) \psi_n(s),
\label{eq:GTDwf}
\end{eqnarray}
which is the same as that of the quark field.
(To be strict, there can appear an irrelevant $n$-dependent 
global phase factor $e^{i\varphi_n[V]}$, 
according to arbitrariness of the phase in 
the basis $|n \rangle$
\cite{GIS12}.)

Note here that the spectral density $\rho(\lambda)$ 
of the Dirac operator $\Slash D$ relates to chiral symmetry breaking.
For example, the Banks-Casher relation \cite{BC80} shows that 
the quark condensate $\langle\bar qq \rangle$ originates from 
the zero-eigenvalue density $\rho(0)$ 
in the limit of large space-time volume $V_{phys}$ 
and in the chiral limit, i.e., 
$
\langle \bar qq \rangle=-\lim_{m \to 0} \lim_{V_{phys} \to \infty} 
\pi\rho(0).
$
In fact, the low-lying Dirac modes can be regarded as the essential modes 
for spontaneous chiral-symmetry breaking in QCD.

\subsection{Operator formalism in lattice QCD}

In this subsection, we present the operator formalism 
in lattice QCD \cite{S1112,GIS12,IS13}. 
We first introduce the link-variable operator $\hat U_{\pm \mu}$ 
defined by the matrix element of 
\begin{eqnarray}
\langle s |\hat U_{\pm \mu}|s' \rangle 
=U_{\pm \mu}(s)\delta_{s\pm \hat \mu,s'}.
\end{eqnarray}
The Dirac operator and the covariant derivative 
are simply written with the link-variable operator as 
\begin{eqnarray}
\Slash{\hat D}
=\frac{1}{2a}\sum_{\mu=1}^{4} \gamma_\mu (\hat U_\mu-\hat U_{-\mu}),
\qquad 
\hat D_\mu=\frac{1}{2a}(\hat U_\mu-\hat U_{-\mu}).
\label{eq:Dop}
\end{eqnarray}
Also, the Polyakov loop $\langle L_P \rangle$ 
is simply expressed as the functional trace of 
$\hat U_4^{N_t}$,
\begin{eqnarray}
\langle L_P \rangle
=\frac{1}{N_c V} \langle {\rm Tr}_c \{\hat U_4^{N_t}\}\rangle
=\frac{1}{N_c V}\left\langle 
\sum_s {\rm tr}_c \left(\prod_{n=0}^{N_t-1} U_4(s+n\hat t)\right)\right\rangle,
\label{eq:PL}
\end{eqnarray}
with the four-dimensional lattice volume $V \equiv N_s^3 \times N_t$ and 
$\hat t=\hat 4$.
Here, ``${\rm Tr}_c$'' denotes the functional trace 
of ${\rm Tr}_c \equiv \sum_s {\rm tr}_c$ including 
the trace ${\rm tr}_c$ over color index.

The Dirac-mode matrix element of the link-variable operator 
$\hat U_{\mu}$ can be expressed with $\psi_n(s)$ as 
\begin{eqnarray}
\langle m|\hat U_{\mu}|n \rangle=\sum_s\langle m|s \rangle 
\langle s|\hat U_{\mu}|s+\hat \mu \rangle \langle s+\hat \mu|n\rangle
=\sum_s \psi_m^\dagger(s) U_\mu(s)\psi_n(s+\hat \mu).
\end{eqnarray}
Note that the matrix element is gauge invariant \cite{GIS12} 
due to the gauge transformation property (\ref{eq:GTDwf}), 
\begin{eqnarray}
\langle m|\hat U_\mu|n \rangle
&\rightarrow&
\sum_s\psi^\dagger_m(s)V^\dagger(s)\cdot V(s)U_\mu(s)V^\dagger(s+\hat \mu)
\cdot V(s+\hat \mu)\psi_n(s+\hat \mu) \nonumber\\
&=&\sum_s\psi_m^\dagger(s)U_\mu(s)\psi_n(s+\hat \mu)
=\langle m|\hat U_\mu|n\rangle.
\end{eqnarray}
To be strict, an irrelevant $n$-dependent global phase factor can appear 
according to the arbitrariness of the phase in the basis 
$|n \rangle$. However, this phase factor exactly cancels 
as $e^{i\varphi_n} e^{-i\varphi_n}=1$ 
between $|n \rangle$ and $\langle n |$, and does not appear 
for physical quantities such as the Wilson loop and the Polyakov loop 
\cite{GIS12}.

\section{Derivation of an analytical relation 
between the Polyakov loop and Dirac modes 
in temporally odd-number lattice QCD}

In this section, we consider temporally odd-number lattice QCD 
\cite{SDI13,DSI13}, 
with the temporal lattice size $N_t$ being odd, 
as shown in Fig.2. 
Apart from the odd-number $N_t$, all the lattice conditions are ordinary.
In fact, we use an ordinary square lattice 
and the normal (nontwisted) periodic boundary condition 
for the link-variable $U_\mu(s)$ in the temporal direction. 
The spatial lattice size $N_s$ is taken to be larger than $N_t$, 
i.e., $N_s > N_t$. 
Note that, in the continuum limit of $a \rightarrow 0$ and 
$N_t \rightarrow \infty$, 
any number of large $N_t$ gives the same physical result.
Hence, it is no problem to use the odd-number lattice.

\begin{figure}[h]
\begin{center}
\includegraphics[scale=0.3]{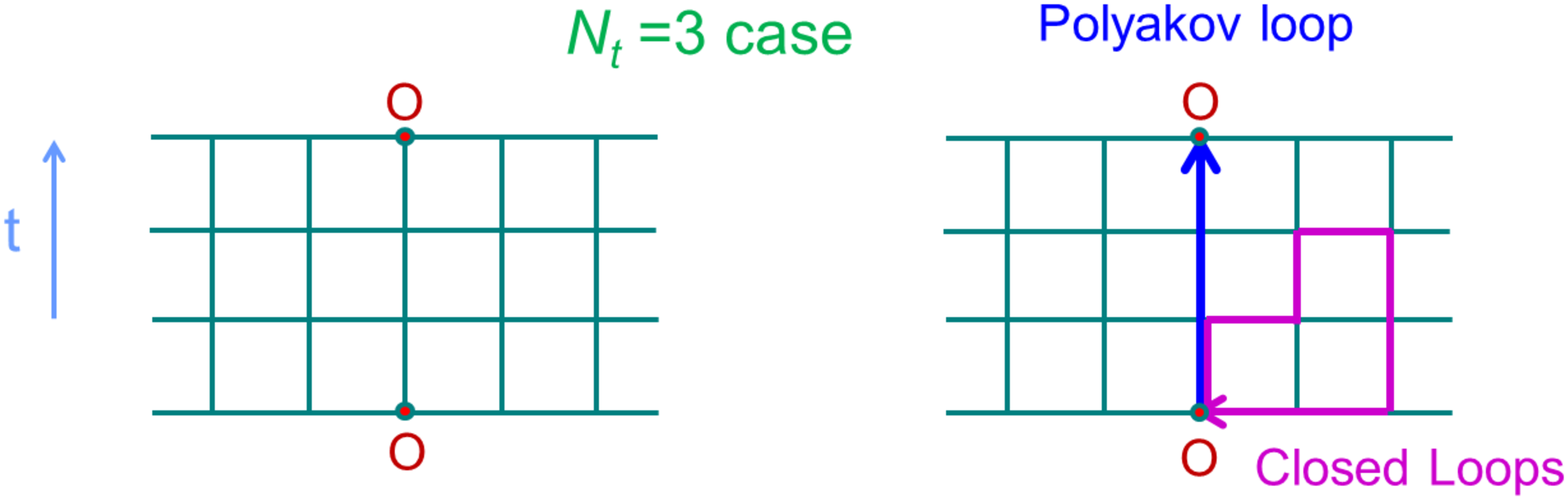}
\caption{
An example of the temporally odd-number lattice ($N_t=3$ case).
Only gauge-invariant quantities such as 
closed loops and the Polyakov loop survive or do not vanish in QCD, 
after taking the expectation value, i.e., the gauge-configuration average.
Geometrically, closed loops have even-number links on the square lattice.
}
\end{center}
\end{figure}

As a general mathematical argument 
of the Elitzur theorem \cite{Rothe12}, 
only gauge-invariant quantities 
such as closed loops and the Polyakov loop survive in QCD.
In fact, all the non-closed lines are gauge-variant 
and their expectation values are zero. 
Note here that any closed loop 
needs even-number link-variables on the square lattice, 
except for the Polyakov loop. (See Fig.2.)

In this temporally odd-number lattice QCD, 
we consider the following functional trace \cite{SDI13,DSI13}:
\begin{eqnarray}
I\equiv {\rm Tr}_{c,\gamma} (\hat{U}_4\hat{\Slash{D}}^{N_t-1}).
\label{eq:FT}
\end{eqnarray}
Here, 
${\rm Tr}_{c,\gamma}\equiv \sum_s {\rm tr}_c 
{\rm tr}_\gamma$ includes 
${\rm tr}_c$ 
and the trace ${\rm tr}_\gamma$ over spinor index.
Its expectation value 
\begin{eqnarray}
 \langle I\rangle=\langle {\rm Tr}_{c,\gamma} (\hat{U}_4\hat{\Slash{D}}^{N_t-1})\rangle 
\label{eq:FTV}
\end{eqnarray}
is obtained as the gauge-configuration average in lattice QCD.
When the volume $V$ is enough large,
one can expect 
$\langle \hat O \rangle \simeq {\rm Tr}~{\hat O}/{\rm Tr}~1$ 
for any operator $\hat O$ even in each gauge configuration.

From Eq.(\ref{eq:Dop}), 
$\hat U_4\!\Slash{\hat D}^{N_t-1}$ 
can be expressed as a sum of products of $N_t$ link-variable operators, 
because the Dirac operator $\Slash{\hat D}$ 
includes one link-variable operator in each direction of $\pm \mu$.
In fact, $\hat U_4\!\!\Slash{\hat D}^{N_t-1}$ 
includes ``many trajectories'' with the total length $N_t$ 
(in the lattice unit) 
on the square lattice, as shown in Fig.3.
Note that all the trajectories with the odd-number length $N_t$ 
cannot form a closed loop 
on the square lattice, and thus give gauge-variant contribution, 
except for the Polyakov loop.

\begin{figure}[h]
\begin{center}
\includegraphics[scale=0.3]{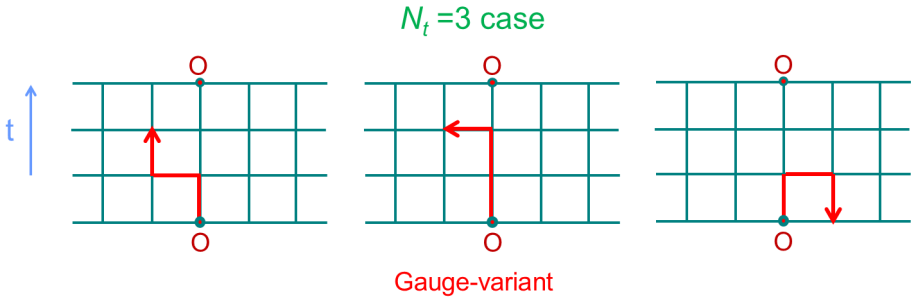}
\caption{
Partial examples of the trajectories stemming from 
$\langle {\rm Tr}_{c,\gamma}(\hat U_4\!\!\Slash{\hat D}^{N_t-1})\rangle$. 
For each trajectory, the total length is $N_t$, and 
the ``first step'' is positive temporal direction 
corresponding to $\hat U_4$.
All the trajectories with the odd-number length $N_t$ 
cannot form a closed loop on the square lattice, 
and therefore they are gauge-variant and give no contribution 
in $\langle {\rm Tr}_{c,\gamma}(\hat U_4\!\Slash{\hat D}^{N_t-1})\rangle$, 
except for the Polyakov loop.
}
\end{center}
\end{figure}

Therefore, among the trajectories stemming from 
$\langle {\rm Tr}_{c,\gamma}(\hat U_4\!\!\Slash{\hat D}^{N_t-1}) \rangle$, 
all the non-loop trajectories are gauge-variant and give no contribution, 
according to the Elitzur theorem \cite{Rothe12}.
Only the exception is the Polyakov loop, as shown in Fig.4. 
(Compare Figs.3 and 4.)
Note here that $\langle {\rm Tr}_{c,\gamma}
(\hat U_4\!\!\Slash{\hat D}^{N_t-1})\rangle$ 
do not include the anti-Polyakov loop $\langle L_P^\dagger \rangle$, 
because the ``first step'' is positive temporal direction 
corresponding to $\hat U_4$.

\begin{figure}[h]
\begin{center}
\includegraphics[scale=0.3]{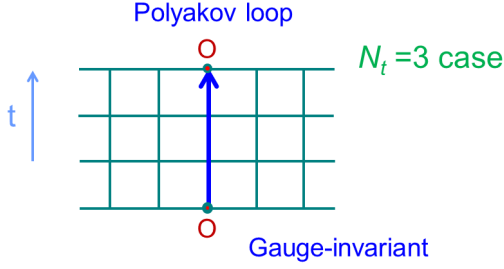}
\caption{
Among the trajectories stemming from 
$\langle {\rm Tr}_{c,\gamma}(\hat U_4\!\Slash{\hat D}^{N_t-1}) \rangle$, 
only the Polyakov-loop ingredient can survive 
as the gauge-invariant quantity. 
Here, $\langle {\rm Tr}_{c,\gamma}(\hat U_4\!\Slash{\hat D}^{N_t-1}) \rangle$ 
does not include $\langle L_P^\dagger \rangle$, 
because of the first factor $\hat U_4$.
}
\end{center}
\end{figure}

In this way, only the Polyakov-loop ingredient can survive 
as the gauge-invariant quantity in the functional trace 
$\langle I \rangle
=\langle{\rm Tr}_{c,\gamma}(\hat U_4\!\!\Slash{\hat D}^{N_t-1})\rangle$, and 
$\langle I \rangle$ is proportional to the Polyakov loop $\langle L_P \rangle$.

Actually, we can mathematically derive the following relation:
\begin{eqnarray}
\langle I\rangle
&=&\langle {\rm Tr}_{c,\gamma} (\hat U_4 \hat{\Slash D}^{N_t-1}) \rangle
\nonumber \\
&=&\langle {\rm Tr}_{c,\gamma} \{\hat U_4 (\gamma_4 \hat D_4)^{N_t-1}\} \rangle
\quad \qquad \qquad ~~~~~{\rm 
(
\raisebox{1.2ex}{.}\raisebox{.2ex}{.}\raisebox{1.2ex}{.} 
~only~gauge\hbox{-}invariant~terms~survive)} 
\nonumber \\
&=&4\langle {\rm Tr}_{c} (\hat U_4 \hat D_4^{N_t-1}) \rangle
\quad \qquad \qquad \qquad ~~~~~~(
\raisebox{1.2ex}{.}\raisebox{.2ex}{.}\raisebox{1.2ex}{.} 
~\gamma_4^{N_t-1}={1}, 
~{\rm tr}_\gamma {1}=4) 
\nonumber \\
&=&\frac{4}{(2a)^{N_t-1}}
\langle {\rm Tr}_{c} \{\hat U_4 (\hat U_4-\hat U_{-4})^{N_t-1}\} \rangle
\quad ~~(
\raisebox{1.2ex}{.}\raisebox{.2ex}{.}\raisebox{1.2ex}{.} 
~\hat D_4=\frac{1}{2a}(\hat U_4-\hat U_{-4}))
\nonumber \\
&=&\frac{4}{(2a)^{N_t-1}} \langle {\rm Tr}_{c} \{ \hat U_4^{N_t} \}\rangle
\qquad \qquad \qquad ~~~~~{\rm 
(\raisebox{1.2ex}{.}\raisebox{.2ex}{.}\raisebox{1.2ex}{.} 
~only~gauge\hbox{-}invariant~terms~survive)} 
\nonumber \\
&=&\frac{12V}{(2a)^{N_t-1}}\langle L_P \rangle. 
\label{eq:FTdetail}
\end{eqnarray}
Thus, we obtain the relation between 
$\langle I\rangle = \langle {\rm Tr}_{c,\gamma}
 (\hat U_4 \hat{\Slash D}^{N_t-1}) \rangle$ 
and the Polyakov loop $\langle L_P \rangle$: 
\begin{eqnarray}
\langle I\rangle
=\langle {\rm Tr}_{c,\gamma} (\hat U_4 \hat{\Slash D}^{N_t-1}) \rangle
=\frac{12V}{(2a)^{N_t-1}}\langle L_P \rangle. 
\label{eq:FTtoPL}
\end{eqnarray}

On the other hand, the functional trace in Eq.(\ref{eq:FTV}) 
can be calculated with the complete set of 
the Dirac-mode basis $|n\rangle$ satisfying $\sum_n |n\rangle \langle n|=1$, 
and we find the Dirac-mode representation of 
\begin{eqnarray}
 \langle I\rangle=\sum_n\langle n|\hat{U}_4\Slash{\hat{D}}^{N_t-1}|n\rangle
=i^{N_t-1}\sum_n\lambda_n^{N_t-1}\langle n|\hat{U}_4| n \rangle. 
\label{eq:FTtoD}
\end{eqnarray}
By combing Eqs.(\ref{eq:FTtoPL}) and (\ref{eq:FTtoD}), 
we obtain the analytical 
relation between the Polyakov loop $ \langle L_P \rangle$ 
and the Dirac eigenvalues $i\lambda_n$ in QCD: 
\begin{eqnarray}
\langle L_P \rangle=\frac{(2ai)^{N_t-1}}{12V}
\sum_n\lambda_n^{N_t-1}\langle n|\hat{U}_4| n \rangle.
\label{eq:PLvsD}
\end{eqnarray}
This is a Dirac spectral representation of the Polyakov loop, 
and is mathematically valid on the temporally odd-number lattice 
in both confined and deconfined phases. 
Based on Eq.(\ref{eq:PLvsD}), 
we can investigate each Dirac-mode contribution 
to the Polyakov loop individually, e.g., 
by evaluating each contribution specified by $n$ 
numerically in lattice QCD.
In particular, by paying attention to low-lying Dirac modes 
in Eq.(\ref{eq:PLvsD}), the relation between confinement 
and chiral symmetry breaking can be discussed in QCD.

\section{Discussions and concluding remarks}

Finally, we discuss the physical meaning of Eq.(\ref{eq:PLvsD}). 
As a remarkable fact, because of the factor $\lambda_n^{N_t -1}$, 
the contribution from 
low-lying Dirac-modes with $|\lambda_n|\simeq 0$ 
is negligibly small in the Dirac spectral sum of RHS in Eq.(\ref{eq:PLvsD}),
compared to the other Dirac-mode contribution. 
In fact, 
{\it the low-lying Dirac modes have quite small contribution 
to the Polyakov loop}, regardless of confined or deconfined phase.

This result is consistent with our previous numerical 
lattice result that confinement properties are almost unchanged by 
removing low-lying Dirac modes from the QCD vacuum \cite{S1112,GIS12,IS13}.

Here, we give several meaningful comments on the relation (\ref{eq:PLvsD}) 
in order.
\begin{enumerate}
\item
Equation (\ref{eq:PLvsD}) is a manifestly gauge-invariant relation. 
Actually, the matrix element $\langle n |\hat U_4|n\rangle$ 
can be expressed with 
the Dirac eigenfunction $\psi_n(s)$ and 
the temporal link-variable $U_4(s)$ as 
\begin{eqnarray}
\langle n |\hat U_4|n\rangle =
\sum_s \langle n |s \rangle \langle s 
|\hat U_4| s+\hat t \rangle \langle s+\hat t|n\rangle
=\sum_s \psi_n^\dagger (s)U_4(s) \psi_n(s+\hat t),
\end{eqnarray}
and each term $\psi_n^\dagger (s)U_4(s) \psi_n(s+\hat t)$ 
is manifestly gauge invariant, due to 
the gauge transformation property (\ref{eq:GTDwf}).
[Global phase factors also cancel exactly 
as $e^{-i\varphi_n}e^{i\varphi_n}=1$ 
between $\langle n|$ and $|n \rangle$.] 
\item
In RHS of Eq.(\ref{eq:PLvsD}), 
there is no cancellation between chiral-pair Dirac eigen-states, 
$|n \rangle$ and $\gamma_5|n \rangle$, because $(N_t-1)$ is even, i.e., 
$(-\lambda_n)^{N_t-1}=\lambda_n^{N_t-1}$, and  
$\langle n |\gamma_5 \hat U_4 \gamma_5|n\rangle
=\langle n |\hat U_4|n\rangle$. 
\item
Even in the presence of a possible 
multiplicative renormalization factor for the Polyakov loop,
the contribution from the low-lying Dirac modes (or 
the small $|\lambda_n|$ region) is relatively negligible, 
compared to other Dirac-mode contribution 
in the sum of RHS in Eq.(\ref{eq:PLvsD}). 
\item
For the arbitrary color number $N_c$, 
Eq.(\ref{eq:PLvsD}) is true and applicable in the SU($N_c$) gauge theory.
\item
If RHS in Eq.(\ref{eq:PLvsD}) {\it were} not a sum but a product, 
low-lying Dirac modes (or the small $|\lambda_n|$ region) 
should have given an important contribution 
to the Polyakov loop as a crucial reduction factor of $\lambda_n^{N_t-1}$. 
In the sum, however, the contribution ($\propto \lambda_n^{N_t-1}$) 
from the small $|\lambda_n|$ region 
is negligible. 
\item
Even if $\langle n |\hat U_4|n\rangle$ behaves as 
$\delta(\lambda)$, the factor $\lambda_n^{N_t-1}$ is still crucial 
in Eq.(\ref{eq:PLvsD}), 
because of $\lambda \delta(\lambda)=0$. 
\item
The relation (\ref{eq:PLvsD}) is correct regardless of 
presence or absence of dynamical quarks, 
although the dynamical quark effect appears in $\langle L_P\rangle$, 
the Dirac eigenvalue distribution $\rho(\lambda)$ and 
$\langle n |\hat U_4|n\rangle$.
\item
The relation (\ref{eq:PLvsD}) is correct also 
at finite density and finite temperature.
\item
Equation (\ref{eq:PLvsD}) obtained on the odd-number lattice 
is correct in the continuum limit of $a \rightarrow 0$ 
and $N_t \rightarrow \infty$, since 
any number of large $N_t$ gives the same physical result.
\end{enumerate}

Most of the above arguments can be numerically investigated by 
lattice QCD calculations. 
Using actual lattice QCD calculations at the quenched level, 
we numerically confirm the analytical relation (\ref{eq:PLvsD}), 
non-zero finiteness of $\langle n|\hat U_4|n\rangle$ for each Dirac mode, 
and the negligibly small contribution of 
low-lying Dirac modes to the Polyakov loop, 
in both confined and deconfined phases \cite{SDI13,DSI13}. 
(Although we numerically find an interesting drastic change of 
the behavior of $\langle n|\hat U_4|n\rangle$ 
between confined and deconfined phases, 
we find also tiny contribution of 
low-lying Dirac modes to the Polyakov loop.)

From the analytical relation (\ref{eq:PLvsD}) and the numerical confirmation, 
we conclude that low-lying Dirac-modes 
have quite small contribution to the Polyakov loop, 
and are not essential for confinement, 
while these modes are essential for chiral symmetry breaking.
This conclusion indicates no direct one-to-one correspondence between 
confinement and chiral symmetry breaking in QCD.

It is interesting to compare with 
other lattice result on importance of 
infrared gluons to confinement: 
confinement originates 
from the low-momentum gluons below 1.5GeV in Landau gauge \cite{YS0809}.
Also, some independence 
between confinement and chiral symmetry breaking 
may lead to richer phase structure in QCD, e.g., 
difference of phase transition points 
between deconfinement and chiral restoration 
in strong electro-magnetic fields, 
due to their nontrivial effect on chiral symmetry \cite{ST9193}.

\section*{Acknowledgements}
H.S. thanks Prof. K. Redlich and Dr. C. Sasaki for useful discussions. 
H.S. and T.I. are supported in part by the Grant for Scientific Research 
[(C) No.23540306, E01:21105006, No.21674002] 
from the Ministry of Education, Science and Technology of Japan.


\begin{thebibliography}{99} 

\bibitem{NJL61} 
Y.~Nambu and G.~Jona-Lasinio, 
Phys. Rev. {\bf 122}, 345 (1961); 
Phys. Rev. {\bf 124}, 246 (1961).

\bibitem{SST95}
H.~Suganuma, S.~Sasaki and H.~Toki, 
Nucl. Phys. {\bf B435}, 207 (1995). \\
H.~Suganuma, S.~Sasaki,  H.~Toki and H.~Ichie,
Prog. Theor. Phys. Suppl. {\bf 120}, 57 (1995). 

\bibitem{M95W95}
O.~Miyamura, Phys. Lett. {\bf B353}, 91 (1995); 
R.M.~Woloshyn, Phys. Rev. {\bf D51}, 6411 (1995).

\bibitem{G06BGH07}
C.~Gattringer, Phys. Rev. Lett. {\bf 97}, 032003 (2006). \\
F.~Bruckmann, C.~Gattringer and C.~Hagen, 
Phys. Lett. {\bf B647}, 56 (2007).

\bibitem{SWL08}
F.~Synatschke, A.~Wipf and K.~Langfeld, 
Phys. Rev. {\bf D77}, 114018 (2008). 

\bibitem{S1112}
H.~Suganuma, S.~Gongyo, T.~Iritani 
and A.~Yamamoto, PoS ({\bf QCD-TNT-II}), 044 (2011). \\
H.~Suganuma, S.Gongyo and T.Iritani, 
PoS ({\bf Lattice 2012}), 217 (2012). 

\bibitem{GIS12}
S.~Gongyo, T.~Iritani and H.~Suganuma, 
Phys. Rev. {\bf D86}, 034510 (2012).

\bibitem{IS13}
T.~Iritani and H.~Suganuma, 
arXiv:1305.4049[hep-lat]; 
T.~Iritani, S.~Gongyo and H.~Suganuma, 
PoS~({\bf Lattice 2012}), 218 (2012); 
PoS ({\bf Confinement X}), 053 (2013).

\bibitem{SDI13}
H.~Suganuma, T.M. Doi and T.~Iritani, 
PoS~({\bf Lattice 2013}), 374 (2013).

\bibitem{DSI13}
T.M. Doi, H.~Suganuma and T.~Iritani, 
PoS~({\bf Lattice 2013}), 375 (2013).

\bibitem{Rothe12} 
H.-J.~Rothe, {\it Lattice Gauge Theories}, 4th edition, 
(World Scientific, 2012) and its references.

\bibitem{LFKRS13}
P.M. Lo, B. Friman, O. Kaczmarek, K. Redlich and C. Sasaki, 
Phys. Rev. {\bf D88}, 014506 (2013); Phys. Rev. {\bf D88}, 074502 (2013).

\bibitem{BC80}
T.~Banks and A.~Casher, Nucl. Phys. {\bf B169}, 103 (1980). 

\bibitem{K02}
F. Karsch, Lect. Notes Phys. {\bf 583}, 209 (2002), 
and its references.

\bibitem{AFKS06}
Y. Aoki, Z. Fodor, S.D. Katz and K.K. Szabo, 
Phys. Lett. {\bf B643}, 46 (2006).

\bibitem{KSW87} 
A.S.~Kronfeld, G.~Schierholz and U.-J.~Wiese,
Nucl. Phys. {\bf B293}, 461 (1987).

\bibitem{SNW94}
J.D.~Stack, S.D.~Neiman and R.J.~Wensley, 
Phys. Rev. {\bf D50}, 3399 (1994). 

\bibitem{AS99} 
K.~Amemiya and H.~Suganuma, 
Phys. Rev. {\bf D60}, 114509 (1999).\\
S.~Gongyo, T.~Iritani and H.~Suganuma, 
Phys. Rev. {\bf D86}, 094018 (2012).\\
S. Gongyo and H. Suganuma, 
Phys. Rev. {\bf D87}, 074506 (2013). 

\bibitem{IS9900}
H. Ichie and H. Suganuma, Nucl. Phys. {\bf B574}, 70 (2000); 
Nucl. Phys. {\bf B548}, 365 (1999).

\bibitem{N74tH81}
Y.~Nambu, Phys. Rev. {\bf D10}, 4262 (1974); 
G.~'t Hooft, Nucl. Phys. {\bf B190}, 455 (1981).

\bibitem{STSM95}
H.~Suganuma, A.~Tanaka, S.~Sasaki and O.~Miyamura, 
Nucl. Phys. Proc. Suppl. {\bf 47}, 302 (1996).

\bibitem{LS11}
C.B.~Lang and M.~Schrock, 
Phys. Rev. {\bf D84}, 087704 (2011).\\
L.Ya~Glozman, C.B.~Lang and M.~Schrock, 
Phys. Rev. {\bf D86}, 014507 (2012).

\bibitem{YS0809} 
A.Yamamoto, H.Suganuma, 
Phys.Rev.Lett.{\bf 101}, 241601 (2008); 
Phys.Rev.{\bf D79}, 054504 (2009).

\bibitem{ST9193}
H.~Suganuma and T.~Tatsumi, Ann. Phys. {\bf 208}, 470 (1991); 
Prog. Theor. Phys. {\bf 90}, 379 (1993).


\end{thebibliography}
\end{document}